\documentclass[aps,prl,twocolumn,amsmath,amssymb,
superscriptaddress,floatfix,showpacs]{revtex4}
\usepackage[dvips]{graphicx}
\begin{document}
\title{A metamagnetic critical point in a three dimensional frustrated antiferromagnet}
 \author{Nic Shannon}
\affiliation{
Max-Planck-Institut f\"ur Physik komplexer Systeme, 
N\"uthnitzer Str. 38, 01187 Dresden, Germany
}
\affiliation{H. H. Wills Physics Laboratory, University of Bristol,  Tyndall Ave, BS8-1TL, UK.}
\author{Karlo Penc}
\affiliation{
Research Institute  for  Solid State  Physics   and
Optics, H--1525 Budapest, P.O.B.  49, Hungary}
\author{Yukitoshi Motome}
 \affiliation{RIKEN (The Institute of Physical and Chemical Research), Wako, Saitama
351--0198, Japan}
\date{\today}
\begin{abstract}   
    The competition between different forms of order is central to the problem of strong correlation.   
    This is particularly true of frustrated systems, which frequently 
    exist at or near to a zero--temperature critical point.
    Here we show that a state with a half--magnetization plateau 
    but no long range order can arise when a three dimensional frustrated 
    antiferromagnet is tuned to a critical point bordering a metamagnetic state.   
    We use classical Monte Carlo simulation and low--temperature expansion techniques 
    to accurately characterize this ``spin pseudogap'' state, and show how its properties 
    relate to those of the critical point.   
    Our results provide 
    an example of  three dimensional spin 
    model which can be used to study the relationship between gap and ``pseudogap'', 
    --- i.e. long range and preformed local order --- near a metamagnetic critical point.   
\end{abstract}
\maketitle

In recent years the concept of ``quantum criticality'' has become
central to efforts to understand correlated electron systems.
This rapidly growing body of work rests on the simple idea that where there 
is a zero--temperature phase transition between
two different ordered phases, the finite temperature properties of 
the paramagnetic phase are controlled by the  
critical point which separates them. 
In order to test this hypothesis experimentally, it is necessary to
identify systems with a suitable control parameter which
can be used to tune through the critical point.   
Mechanical pressure, chemical pressure, doping and magnetic field
have all been used successfully to this end.   
Two widely discussed examples are the underdoped cuprate
superconductors, whose ``pseudogap phase'' has been suggested to
originate in a quantum critical point as a function of doping, 
and Sr$_3$Ru$_2$O$_7$ where non--Fermi liquid behavior appears to be 
associated with a metamagnetic transition with strongly suppressed 
$T_{\rm c}$~\cite{andy}.

\begin{figure}[tb]
  \centering
  \includegraphics[width=7truecm]{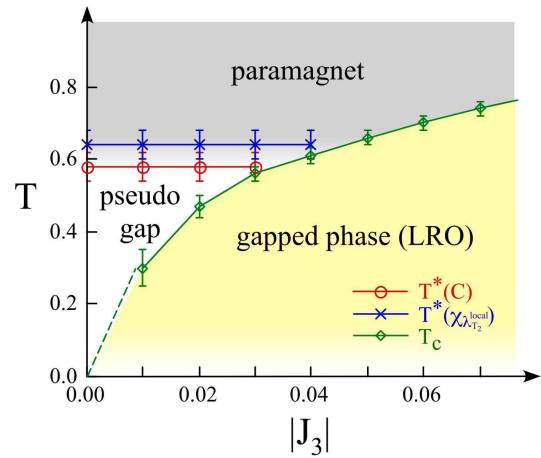}
  \caption{(Color online) 
  Phase diagram for the classical Heisenberg antiferromagnet on a pyrochlore 
  lattice~(\protect\ref{eq:Hb})  in applied magnetic field $h=4$, with additional
  biquadratic interactions $b=0.6$.
  The transition temperature $T_{\rm c}$ associated with the gapped, ordered, half--magnetization plateau state
  vanishes as the strength of ferromagnetic 
  third--neighbour interactions $J_3 \to 0$.   A ``spin--pseudogap'' phase exhibiting a
  half--magnetization plateau but no long--range magnetic order 
  exists above this critical point below a crossover temperature $T^*$. 
  Solid lines are guides for Monte Carlo data, and the dashed line shows 
  $T_{\rm c}$ predicted by low--temperature expansion.
\label{fig:J3}}
\end{figure}

Theoretical attempts to understand these phenomena have largely concentrated
on renormalization group analysis of phenomenological field 
theories~\cite{subir}.  However it is also interesting to ask what happens in 
microscopic 
models, especially where these are accessible to a
variety of different approaches.   Can we construct concrete examples
of systems with a zero--temperature critical point~?   What are the
nature of the correlations at this point, and in the paramagnetic 
phase connected to it~?  
What does a ``pseudogap'' look like at a microscopic level~?

In this Letter we show how a state with a gap to spin excitations 
--- a ``spin pseudogap'' --- 
but no long--range magnetic order, 
can arise near a metamagnetic critical point 
in a microscopic model.
Our main results are summarized in the phase diagram Fig.~\ref{fig:J3}.
The model which we use to explore these ideas is the
antiferromagnetic (AF) Heisenberg model on the 
highly frustrated pyrochlore lattice, in applied magnetic field ${\bf h}$:
\begin{eqnarray}
     \mathcal{H} &=& J \sum_{\langle i,j \rangle} \big[ {\bf S}_i \cdot {\bf S}_j
       - b ({\bf S}_i \cdot {\bf S}_j)^2 \big]
	   \nonumber \\
       &+& J_3 \sum_{\langle i,j \rangle''} {\bf S}_i \cdot {\bf S}_j 
       -  h \sum_i  S^z_i. 
              \label{eq:Hb}
\end{eqnarray}
In order to make the problem accessible to large scale simulation,
we consider the classical limit of the problem $S = |{\bf S}| \to \infty$.
The dominant effect of quantum fluctuations in frustrated systems
is known to be a tendency towards collinearity, and we characterize this
through an effective biquadratic interaction $b$~\cite{oldpaper}.  
An additional 
third--neighbour interaction $J_3$ 
is used to tune the system away from a critical point at $J_3=0$.
The model~(\ref{eq:Hb}) was recently shown to offer an explanation of 
the broad half--magnetization plateaux observed in the spinel 
oxides CdCr$_2$O$_4$ and HgCr$_2$O$_4$~\cite{penc04-PRL,ueda05-PRL}.
In these materials, strong 
effective biquadratic interactions 
$b \sim 0.1$--$0.2$ arise from the coupling of spins to the lattice. 

\begin{figure}[tb]
  \centering
  \includegraphics[width=8truecm]{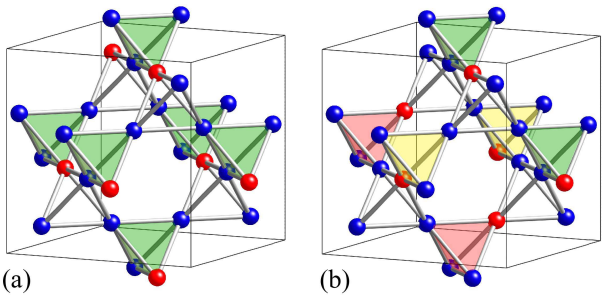}
  \caption{(Color online) 
  Half--magnetization plateau states ($uuud$ states) on a pyrochlore lattice 
  with exactly  three up (black) and one down (white) spins per tetrahedron.
  (a) $uuud$ state with long--range four--sublattice order, as considered in 
  \protect\cite{penc04-PRL}.  
  (b) A schematic picture of 
  $uuud$ state with no long range order of the type
  found at the critical point.
\label{fig:liquid-solid2.eps}}
\end{figure}

For simplicity, we focus below on the limiting case of ferromagnetic (FM) 
$J_3 \to  0^-$.   For FM $J_3$, the model~(\ref{eq:Hb}) exhibits 
${\bf q} = {\bf 0}$ four--sublattice 
long--range magnetic order at low temperatures~\cite{penc04-PRL}.
The most dramatic feature of its zero--temperature phase diagram 
is a first order transition in applied field  into a metamagnetic ``plateau'' 
state with 
$m =1/2$ --- exactly half the saturation magnetization --- protected by a gap to the lowest lying spin excitation.
This state is of the type illustrated in Fig.~\ref{fig:liquid-solid2.eps}(a).  
We have performed extensive Monte Carlo (MC) simulations of
Eq.~(\ref{eq:Hb}) at finite temperatures~\cite{yuki}, 
complemented by classical low--temperature expansions about both ordered and
disordered states. 
These confirm that the phase diagram given in~\cite{penc04-PRL}
remains valid up to a transition temperature $T_{\rm c}$ 
whose energy scale is set by $|J_3|$.
However for $|J_3| \to 0$, the transition temperature for each of the 
different ordered phases vanishes and the system exists at a critical point.

\begin{figure}[tb]
  \centering
  \includegraphics[width=7truecm]{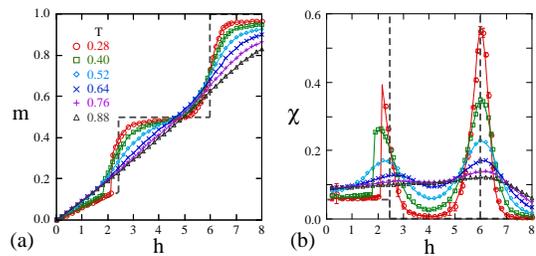}
  \caption{(Color online) 
  Dependence of (a) magnetization $m$ and (b) uniform magnetic 
  susceptibility $\chi$ on magnetic field $h$ for $b=0.6$ and $J_3=0$, 
  showing the 
  existence of the magnetization plateau in the absence of long--range magnetic order.
  Symbols (lines) denote the data for $L=16$ ($L=8$).
  The dashed lines show the $T=0$ results for 
  long--range four--sublattice order. 
\label{fig:plateau}}
\end{figure}

So much for long range order --- what about metamagnetism ?
In Fig.~\ref{fig:plateau} we illustrate  the magnetization process of Eq.~(\ref{eq:Hb}) for $J_3 = 0$.
Data are taken from  MC simulations of Eq.~(\ref{eq:Hb}) using a Metropolis algorithm 
with local spin update~\cite{MCmethod}.  For convenience we assume a (large) value of $b=0.6$, and 
work in units such that $J = S = 1$.   
It is clear from Fig.~\ref{fig:plateau}(a) that 
the magnetization plateau is alive and well --- 
in fact from the magnetization process alone 
it is essentially impossible to distinguish
these results from those for the four--sublattice ordered phase for 
FM $J_3$~\cite{yuki}.  The strong suppression of the magnetic
susceptibility at low temperatures shown in Fig.~\ref{fig:plateau}(b)
suggests the existence of a well 
defined gap, of similar magnitude to that in the nearby ordered phase.
How should we reconcile these results with the vanishing transition temperature
for the ordered plateau state at $J_3 = 0$~?

\begin{figure}[tb]
  \centering
  \includegraphics[width=7truecm]{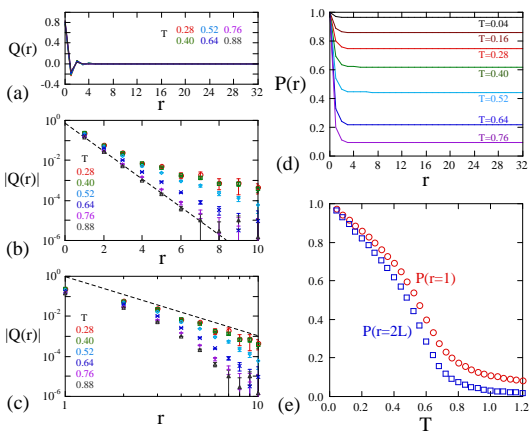}
  \caption{(Color online) Absence of long--range magnetic order in the
  spin--pseudogap state for $b$=0.6, $J_3=0$ and $h$=4.
  (a)---(c) The reduced spin correlation function $Q(r)$, defined 
  by~\protect\mbox{Eq.~(\ref{eq:spin-spin})}, decays exponentially at 
  high temperatures, crossing over to a power law behavior $Q(r) \sim 
  1/r^3$ at low temperatures.   
  (d), (e) However the measure of collinearity $P(r)$, defined by 
  \protect\mbox{Eq.~(\ref{eq:nematic})} {\it does} exhibit long range 
  order at low temperatures.    
  All the results are calculated along [110] chains for $L=16$.
  Distances are measured in units of 
  the nearest-neighbor bond length.
\label{fig:liquid}}
\end{figure}

In Fig.~\ref{fig:liquid} we present MC results for the reduced spin--spin
correlation function
\begin{equation}
    Q({\bf r}_{ij}) = \langle {\bf S}_i \cdot {\bf S}_j \rangle 
      - \langle m \rangle^2,
    \label{eq:spin-spin}
\end{equation}
and the measure of collinearity 
\begin{equation}
    P({\bf r}_{ij}) = \frac{3}{2}\left[ \langle \left( {\bf S}_i \cdot 
    {\bf S}_j \right)^2 \rangle - \frac{1}{3}\right]. 
    \label{eq:nematic}
\end{equation}
As shown in Fig.~\ref{fig:liquid}(a),
spin correlations exhibit a liquid--like structure.  After 
an initial AF oscillation on the scale of a near--neighbour they decay
very rapidly to zero.   Within the range 
accessible to 
MC simulation, the envelope for this decay appears to cross over
smoothly between an exponential decay
at high temperatures and 
a $1/r^3$ power--law behavior at low temperatures, 
as shown in Figs.~\ref{fig:liquid}(b) and (c).   
The collinearity $P({\bf r}_{ij})$,
on the other hand, {\it does} exhibit long range order
--- see Figs.~\ref{fig:liquid}(d) and (e).  

We can understand these results as follows --- 
the biquadratic interaction $b$, which we have introduced
to characterize quantum effects, favors collinear 
states.  The pyrochlore lattice is corner--sharing network of
tetrahedra.  For $h \simeq 4$, 
the $b$ term acts to select $uuud$ states with exactly
three up and one down spins in each tetrahedron and, consequently, $m=1/2$.
The ${\bf q}={\bf 0}$ ordered phase [Fig.~\ref{fig:liquid-solid2.eps}(a)]
is an example of such states.
But there are many more --- in fact there
is a one--to--one mapping between $uuud$ states on a pyrochlore lattice 
and the $\sim 1.3^{N/2}$ dimer coverings of the dual diamond
lattice~\cite{nagler}.
Exactly at the critical point, for $T=0$ and $J_3 = 0$, the system exists in
an equally weighted superposition of all $uuud$ states, 
exhibiting a long--range spin collinearity but no magnetic order.
A schematic picture of this spin--liquid plateau state is shown in 
Fig.~\ref{fig:liquid-solid2.eps}(b).

We can use this insight into the nature of the ground state manifold
at the critical point to construct a low--temperature theory for the
paramagnetic phase connected with it.   Expanding in small
fluctuations about the dimer
manifold, we find a free energy per spin 
\begin{eqnarray}
{\mathcal F} 
&=& J - 3bJ -h 
 - T \ln T + 
 T \langle  \ln |{\bf M}| \rangle/2 \nonumber\\
 && \quad - 
 (T \ln 1.3)/2 + {\mathcal O}(T^2),
\end{eqnarray}
where ${\bf M}$ is the Hamiltonian matrix for harmonic oscillations about 
a given $uuud$ state, and the average $\langle \ldots \rangle$ is
taken in the manifold of all $uuud$ states.   Fluctuation effects, and in 
particular the magnetic susceptibility $\chi(h,T)$, depend
on the model parameters only through the entropy term 
$T\langle \ln |{\bf M}|\rangle/2$.

\begin{figure}[tb]
  \centering
  \includegraphics[width=6truecm]{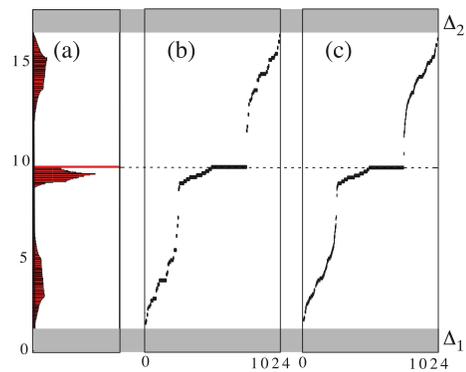}
  \caption{(Color online) 
  (a) DOS of four--sublattice $uuud$ state in thermodynamic limit, showing 
  finite gap and flat band at finite energy.
  (b) Cumulative DOS of four--sublattice $uuud$ state in 1024 spin cluster.
  (c) Cumulative DOS of typical disordered $uuud$ state in 1024 spin cluster.
  In all cases 
  $J_3=0$, $h=4$, and $b=0.6$.
  \label{fig:dos}}
\end{figure}

In the limit $b \to 0$, all $uuud$ states have {\it exactly} the same excitation 
spectrum, with four distinct ``bands'', two of them non--dispersing zero modes.  
Finite values of $b$ lift the degeneracy 
of the spectra of different $uuud$ states, and open a gap $\Delta_1$ to the first spin excitation.
We have performed a MC search within the manifold of $uuud$ configurations 
for finite values of $b$, evaluating the excitation spectrum and $\ln |{\bf M}|$ 
for each state numerically on a finite lattice of $N = 1024$ spins.  
Typical results for the cumulative density of states (DOS) are shown in Fig.~\ref{fig:dos}.
We find that the spectrum is always bounded above and below by the same gaps 
$\Delta_2$ and $\Delta_1$, which are set by the same nodeless excitations.
In the four--sublattice ordered $uuud$ state these reduce to the highest and lowest
energy excitations with ${\bf q} = {\bf 0}$, respectively.  
One of the zero modes survives at finite $b$ as a non--dispersing excitation at finite energy.  
Variation in entropy between different $uuud$ states is negligible compared with the entropy of the 
$uuud$ manifold, and so fluctuations alone cannot drive the system to order.   
The insensitivity of the excitation spectrum to the type of $uuud$ 
state considered explains why the thermodynamic properties of the 
disordered phase for $J_3 = 0$ --- in particular its 
half--magnetization plateau --- are so similar to those of the ordered 
phases for finite $|J_3|$.

We can go further and calculate the asymptotic decay of spin correlations 
in the region of  the critical point.   We do this by constructing an effective 
electrodynamics for the $uuud$ ``dimer'' manifold in close analogy with a 
recent treatment of the $uudd$ ``ice'' manifold~\cite{henleyice}. 
In this approach, the entropy term $T\ln |{\bf M}|/2$ is written in
terms of a course--grained polarization field ${\bf P}$ which measures 
deviations from $uuud$ order.   We find that the spin
correlations in the $uuud$ manifold decay as $1/r^3$, just like those of
the $uudd$ manifold.  This result is in perfect agreement with the results of our
MC simulation at low temperatures --- see Fig.~\ref{fig:liquid}(c). 

Thus the situation in the immediate vicinity of the zero--temperature 
critical point for $J_3=0$ is fairly clear.  A power law behavior of spin correlations
coexists with a finite gap to spin excitations and a well defined plateau in the
magnetization process.    Since this gap is {\it not} associated 
with long range order --- at least in any conventional sense --- we 
adopt the terminology ``spin pseudogap'' to describe this state.

\begin{figure}[tb]
  \centering
  \includegraphics[width=7truecm]{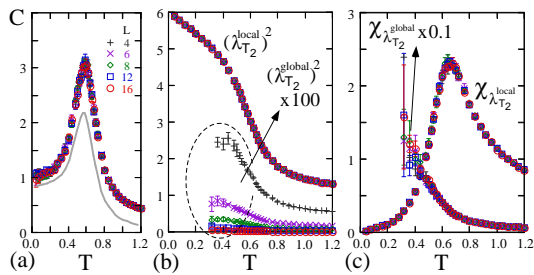}
  \caption{(Color online) 
  Crossover into the spin--pseudogap phase as function of $T$ 
  for $J_3 = 0$, $h=4$ and $b=0.6$ :
  Temperature dependence of (a) the specific heat (the gray 
  line shows the contribution of the 
  biquadratic term alone), (b) the 
  local and global order parameters  $\lambda_{\sf T_2}$ defined by 
  \protect\mbox{Eq.~(\ref{eqn:orderparameters2})}, 
  and (c) the susceptibilities $\chi_{\lambda}$ associated with each of these order 
  parameters.   
\label{fig:crossover}}
\end{figure}

But how does this ``spin pseudogap'' state emerge from the 
paramagnet at high temperatures ? 
And how does it relate to the state for FM $J_3$ with long--range magnetic order~?
To answer these questions, 
we introduce measures of both local and global order, 
based on the relevant irreducible representation of the 
tetrahedral symmetry group~$T_d$ :
\begin{eqnarray}
    \lambda_{\sf T_2}^{\rm local} = \frac{2}{N}\sum_{\rm tetra} 
    {\bf \Lambda}^2_{\sf T_2}, \quad 
    \lambda_{\sf T_2}^{\rm global} = \frac{2}{N}\Big[ \sum_{\rm 
    tetra} 
     {\bf \Lambda}_{\sf T_2} \Big]^2,
     \label{eqn:orderparameters2}
\end{eqnarray}    
where, following~\cite{penc04-PRL}, $ {\bf \Lambda}_{\sf T_2}  
= (\Lambda_{{\sf T_2},1} ,\Lambda_{{\sf T_2},2} ,\Lambda_{{\sf T_2},3}
)$, and $ \Lambda_{{\sf T_2},1} = 
( {\bf S}_2 \cdot {\bf S}_3 
- {\bf S}_1 \cdot {\bf S}_4 )/\sqrt{2}$, etc., are defined in terms of the spins on an
individual tetrahedron. 
We also consider the related susceptibilities, 
$
     \chi_\lambda(T) = \left[ \langle  \lambda^2 \rangle 
   -  \langle  \lambda \rangle^2\right]/T .
$

In Fig.~\ref{fig:crossover} we show MC results for 
the specific heat,  local and global ``order'' parameters $\lambda_{\sf T_2}$
and their related susceptibilities.   
A broad peak in the specific heat is observed at $T \approx b$, 
with most of the entropy coming 
from the selection of collinear states.   At the same time, the
measure of local order $\lambda_{\sf T_2}^{\rm local}$ increases
rapidly, and the related susceptibility $\chi_{\lambda_{\sf T_2}}^{\rm local}$
shows a broad peak.   However none of this signals the onset 
of long--range four--sublattice $uuud$ order of the type found for FM $J_3 < 0$ 
--- the relevant order parameter $\lambda_{\sf T_2}^{\rm global}$ tends 
to zero in the thermodynamic limit, and the related susceptibility
$\chi_{\lambda_{\sf T_2}}^{\rm global}$ appears to diverge only as $T \to 0$.

The peak in $\chi_{\lambda_{\sf T_2}}^{\rm local}$ 
or,
equivalently, the peak in the specific heat $c$ defines a crossover 
temperature $T^*$ 
for entering the spin pseudogap state, with its associated
magnetization plateau.   
The crossover also corresponds to an emergence of the power law behavior
in spin correlations and the rapid development of collinearity
in Fig.~\ref{fig:liquid}.
$T^*$ is set by $b$ --- in fact, we have confirmed that
for $b < J$, it scales nearly linearly with $b$.  
Meanwhile, 
the global order parameter monitors a transition 
for long--range ordered phase for finite $J_3$.
By tracking both $T^*$ and $T_{\rm c}$ as a function of $J_3$, we can map
out the phase boundary of the true, long--range ordered plateau phase,
and the extent of the finite--temperature spin pseudogap ``phase''
associated with the critical point at $J_3 = 0$.   
Results for $b=0.6$ and $h=4$ are shown in Fig.~\ref{fig:J3}.   
In the vicinity of the critical point, 
we can also estimate $T_{\rm c}$ for this (1$^{st}$ order) transition 
within the low--temperature expansion
by comparing 
the change in internal energy with the entropy change going from the gapped to the 
pseudogapped states:   
The prediction is $T_{\rm c} \sim 30|J_3|$, 
the dashed line in Fig.~\ref{fig:J3}.
This phase diagram summarizes the relationship between gap and ``pseudogap'', 
or order and disorder, near a metamagnetic critical point.

For small $|J_3|$ where $T_{\rm c} < T^*$, 
the system exhibits a magnetization
plateau in the absence of long range order.     As the system is 
cooled from the spin--pseudogap state into the ordered phase, 
the preformed local order in ${\bf \Lambda}_{\sf T_2} $ 
becomes global order in $\sum_{\rm tetra} {\bf \Lambda}_{\sf T_2} $.
This provides an intriguing magnetic analogy for the way in which 
preformed superconducting order is believed to emerge from preformed
local cooper pairs in underdoped cuprates.

So far as Cr spinels are concerned, our results suggest that a
material which exhibits a magnetization plateau while maintaining the full
(cubic) symmetry of the pyrochlore lattice, 
need not be magnetically ordered~\cite{mike}.  

In the light of our results it also seems highly probable that the spin--1/2
Heisenberg model on a pyrochlore lattice exhibits a half--magnetization 
plateau because of strong quantum fluctuations.   
Within a spin--wave approximation, our calculations show that the spin gap protecting the 
plateau state is again finite and set by a nodeless wave function, and so is insensitive to 
whether the system is ordered or disordered.

We conclude by noting that it is possible to construct a phase diagram similar to 
Fig.~\ref{fig:J3} for Eq.~(\ref{eq:Hb}) near critical point with $J_3=0$ and $h=0$.   
In this case the competing ground states belong to the $uudd$ ``ice'' 
manifold, and the correlations at the critical point
are of the spin--liquid form described in~\cite{moessner-chalker}.
However there are two important distinctions 
from the case of $h=4$.
Firstly, neither ordered nor disordered $uudd$ phases possesses a gap to spin excitations.   
And secondly, for $h=0$, the collinearity
Eq.~(\ref{eq:nematic}) is a well defined nematic order parameter.

While we have been unable to construct any {\it local} object which can 
serve as order parameter for the spin--pseudogap state for $h=4$, we cannot rule 
out the possibility that $T^*$ signals the onset of some exotic 
non--local order~\cite{wen}.   This remains as a subject for future investigation.

{\it Acknowledgements}
It is our pleasure to acknowledge stimulating discussions with T.~Momoi, 
H.~Takagi, O.~Tchernyshyov, H.~Tsunetsugu and
M.~E.~Zhitomirsky.
This work was supported under a Grant--in--Aid for Scientific Research
(No. 16GS50219) and NAREGI 
from the Ministry of Education, Science, Sports, and Culture of Japan,
Hungarian OTKA T049607 and SFB 463 of the DFG.


\begin{thebibliography}{99}
\bibitem{MCmethod}
We typically perform $10^5$--$10^6$ MC samplings for measurements after 
$10^4$--$10^5$ steps for thermalization.    System sizes are measured in 
units of the the $16$ spin cubic unit cell, so that the total number of spins $N = 16 \times L^3$.  
Because of the frustration inherent in \protect\mbox{Eq.~(\ref{eq:Hb})}, the autocorrelation time 
becomes extremely large at low temperatures.   
Data are shown only where simulations have converged and are known to be ergodic.
\end{thebibliography}
\end{document}